# A novel hybrid and publicly available model for spur gear vibrations based on an efficient dynamic model

Omri Matania[1], Lior Bachar[1], Roee Cohen, Jacob Bortman

**Abstract:**

Dynamic models hold great potential for research and development in signal processing, machine learning, and digital twin algorithms for diagnosing rotating machinery. Various studies have suggested dynamic models of gears, employing many model approaches. However, there is currently a lack of a computationally efficient and publicly accessible model that accurately represents real-world data. In this study, we propose a novel hybrid model that integrates a realistic and efficiently validated dynamic model of spur gear vibrations with an enhancement process aimed at bridging the gap between simulation and reality. This process minimizes discrepancies between features extracted from simulated and measured data through fine-tuning of the model hyperparameters. The effectiveness of this hybrid model is demonstrated across numerous test apparatuses, encompassing several types of faults, severities, and speeds. The new hybrid model, inclusive of an upgraded dynamic model, generates data swiftly within seconds and is made publicly available with a user-friendly application programming interface and a detailed user manual. The novel suggested hybrid model has great potential to enhance future research in model-based studies, including machine learning, signal processing, and digital twin approaches.

The user manual can be found in the following link:
https://github.com/PHM-BGU/public_dynamic_model_for_gear_vibrations

---

[1] Equally contributed authors

## 1. Introduction:

Dynamic models are pivotal in the field of prognosis and health management of rotating machinery [1] as they facilitate the investigation of phenomena expressed in the vibration signal associated with the component's physics [2]. Moreover, they support the development of innovative physics-based algorithms [3,4]. In the current era of machine learning [5–7] and digital twins [8,9], these models possess significant advantages for data-driven algorithms [3]. They enable the generation of diverse data representing various scenarios that are challenging to replicate in real systems and test apparatuses. Additionally, in digital twin applications, these models allow simulation of the real twin's environment, thereby contributing to the decision-making process. This study is focused on gear vibrations, which are widely used for fault diagnosis of gears.

Over the years, numerous dynamic lumped parameter models for gear vibrations have been proposed, establishing a well-defined modeling scheme. Notable literature reviews in this field, conducted by Liang et al. [10] and Mohammed et al. [11], provide comprehensive insights. Table 1 summarizes progress in key aspects of dynamic modeling of gears, encompassing the assumed number of degrees of freedom, methods for calculating time-variant gear mesh stiffness, gear types, modeled fault types, and complex phenomena resulting from system imperfections.

Table 1 - An overview of key aspects in dynamic models of gear vibrations alongside representative literature

| Degrees of Freedom | Gear Mesh Stiffness | Gear Type | Gear Faults | Imperfections |
|---|---|---|---|---|
| 1 [12] | potential-energy [13] | spur [14–18] | pitting [19,20] | backlash [21,22] |
| 3 [14] | square wave [19] | helical [23,24] | cracks [19,25] | transmission error [12] |
| 6 [26] | finite element [27,28] | planetary [19] | wear [23,29] | surface roughness [30,31] |
| 12 [30] | | worm [32] | breakage [33] | lubrication [34] eccentricity [35] |

Three key findings regarding the use of dynamic modeling can be deduced from the literature: (1) A notable limitation is the absence of shared dynamic models with public access [2,36], which hampers their practical application. It is worth highlighting that many algorithmic researchers, particularly those specializing in machine learning, are not necessarily model developers, and vice versa. However, algorithmic researchers can greatly benefit from the availability of dynamic models [3,36]. Consequently, when beneficiaries of these models do not have appropriate access to them, the scientific community as a whole suffers a loss. (2) Dynamic models necessitate complex computations that consume significant running time, yet this issue is seldom addressed in the literature [2,36], possibly due to the challenges associated with it or a lack of prioritization. However, the significance of this challenge becomes apparent in applications that rely on large volumes of data, such as machine learning [5,37]. In the context of digital twins, faster running times enable the examination of numerous cases to inform the decision-making process. Moreover, the ability to swiftly generate data facilitates the exploration of various parameters, and allows for efficient analysis of physics-based algorithms [3]. (3) Every model is underlain by a set of assumptions that results in differences between the simulated and real signals [36]. There is a lack of published work on enhancing simulated signals to reduce these discrepancies. However, for most algorithms, the requirement for a good resemblance between the simulated and measured signals is crucial.

This study addresses the above-mentioned challenges by proposing a hybrid model to generate realistic simulations of spur gear vibrations. The hybrid model comprises an upgraded dynamic model based on former work of Dadon et al. [30], and a model enhancement process. The main novelties of this study are as follows:

1. Contribution of a readily accessible and user-friendly shared dynamic model for spur gear vibrations that can be activated by individuals without specialized expertise. To the best of the authors knowledge, this is the first evidence of sharing a complete dynamic model with the research community. The model is packaged with a well-structured application programming interface and includes a detailed user manual.
2. The upgraded dynamic model generates data swiftly within seconds to tens of seconds on regular personal computers. The improved running time is the result of optimization through careful examination of each component within the model.
3. Introducing a novel hybrid modeling approach to minimize discrepancies between simulated and real measured signals through simulation tuning. Three enhancement procedures are introduced, namely modifications to the signal duration, fault-to-harmonics ratio, and signal-to-noise ratio. The effectiveness of this approach is demonstrated on real test apparatuses featuring spur gears with common fault types.

The structure of this work is organized in the following order. Section 2 and Section 3 introduce the dynamic model for spur gears and the model application programming interface. Section 4 presents the process of computational time optimization. Section 5 introduces the novel hybrid approach for simulation enhancement, demonstrated on experimental measured data. Section 6 summarizes and concludes the presented work.

## 2. The dynamic model:

Dynamic models for generating simulated vibration data are a strong tool for researchers in the field of health management of rotating machinery [3,36]. This work contributes to the field an experimentally validated 3D dynamic model with 13 degrees of freedom for simulating vibrations of spur gear. The model generates the vibration signal by constructing and solving the non-linear Euler-Lagrange equations of motion as illustrated in Equation 1.

$$M\ddot{u} + C\dot{u} + K(u) \cdot u = F_{ex} \qquad (1)$$

Where $u$ is the vector of generalized coordinates, $\dot{u}$ and $\ddot{u}$ are, respectively, the velocities and accelerations, the square matrices $M, C, K(u)$ correspond to the mass, damping, and time-variable stiffness matrices, and $F_{ex}$ denotes the vector of external forces.

The general flow of the model is depicted in Figure 1 and can be described as an eight-steps process:
- Stage 1 – generating structures with model parameters based on the desired spur gear transmission, input speed, output load, tooth surface roughness, and predefined default parameters. The user has a flexibility to modify any of the default parameters such as pressure angle, fault type, material, sampling rate, initial conditions, etc.
- Stage 2 – generating the profile of healthy teeth and determining the properties of the contact line, including the contact ratio and the initial contact point along the involute.
- Stage 3 – incorporating tooth profile errors resulting from manufacturing errors associated with the desired surface quality, adhering to the DIN-3962 standard, and based on principles introduced in Ref. [31].
- Stage 4 – creating a structure that includes information on the tooth faults based on the desired fault and dimensions.
- Stage 5 – calculating the gear mesh stiffness using beam theory principles and contact analysis [38].
- Stage 6 – determining the components of the Euler-Lagrange equations from Equation 1, including the mass, damping and stiffness matrices, as well as the external forces vector.
- Stage 7 – numerically solving the Euler-Lagrange equations using the Newmark [39] and Newton-Raphson methods.
- Stage 8 – generating the simulated vibration signal based on the numerical solution obtained from the previous stage.

For a more in-depth understanding of the model, its application, and the modeling methods employed, please consult the user manual [40].

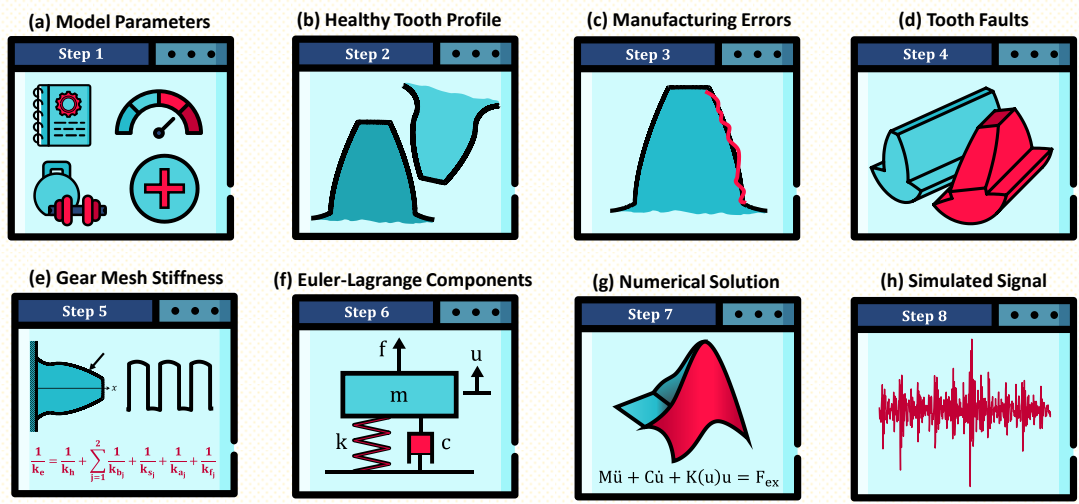

*Figure 1 - Building blocks of the dynamic model, arranged by steps.*

## 3. Model Application Programming Interface (API):

The accessibility of a strong dynamic model to the research community holds great significance for a wide range of applications, including machine learning [5,41], digital twins [8], generating realistic simulations [2,36], and signal processing research [42,43]. To the best of the author's knowledge, there is currently no online dynamic model available that enables users to effectively leverage models and swiftly generate realistic simulations while benefiting from a comprehensive application programming interface (API).

This study places significant importance on the contribution of such a model and provides a well-structured API for a realistic dynamic model that simulates spur gear vibrations. The code library associated with this model is freely accessible online for non-profit academic purposes, such as research and courses, provided proper citation of the current study.

The model is activated using a function called *'generate_gear_simulated_data'*. This function gets information about the desired spur gear transmission, operational conditions, and tooth surface roughness. Users have the flexibility to change default parameters, such as fault type. The library provides various example scripts for different fault types and special cases. For instance, it allows for the generation of multiple signals with the same manufacturing profile errors. A graphic illustration for the model output in case of a tooth breakage fault is depicted in Figure 2.

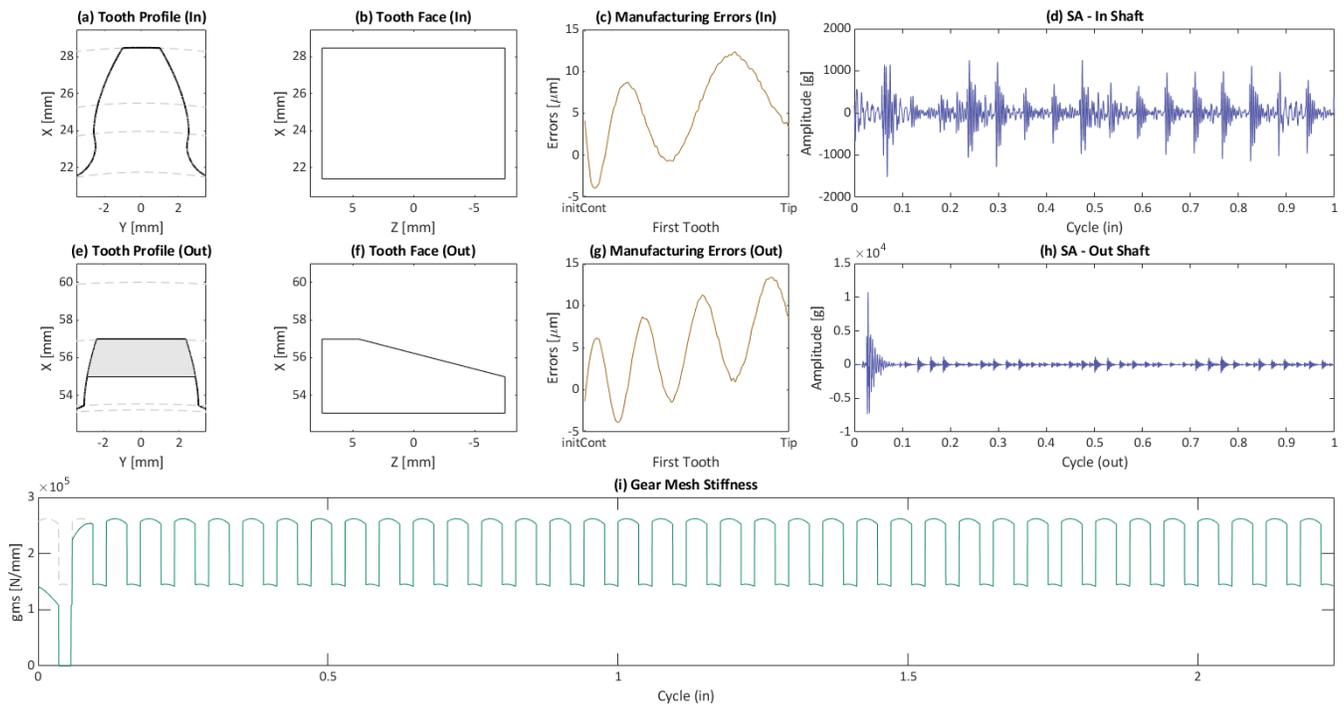

*Figure 2 - An example of the results panel showing tooth breakage fault in the output wheel, including the teeth profile in two views, profile errors, synchronous average signal for each shaft and the gear mesh stiffness.*

## 4. Computational time optimization:

Accelerating computational running time in a model that generates synthetic data for machine and deep learning has significant implications in data-driven technologies like digital twins. By reducing data generation time, researchers can accelerate training and evaluation, leading to quicker iterations and improved productivity. Faster computational running time also enables exploration of larger, diverse datasets, facilitating a comprehensive understanding of complex scenarios, thus optimizing algorithms for more accurate and reliable results.

The dynamic model introduced in Sections 2 and 3 has undergone significant improvements with a focus on accelerating computational time. In addition to conventional methods such as loop avoidance and utilizing vector and matrix multiplications, various computational enhancements were implemented. Three representative examples are given in the paragraphs below:

1. Efficient calculation of the strain energy – the non-linear gear mesh stiffness assists beam theory for calculations. The potential strain energies resulting from axial, bending, and shear stresses are computed for each point along the tooth axis $X_i$ ($i = 1, ..., N$). For instance, Equation 2 demonstrates the strain energy of a healthy tooth pair caused by bending stress at the $i^{th}$ index. To improve computation efficiency, the integral in Equation 2 is divided into three separate integrals, as shown in Equation 3, moving independent coefficients out of the integral. This approach allows for direct calculation of the strain energy without the need for loops, resulting in a complexity of $O(N)$ instead of the original $O(N^2)$ complexity.

$$U_b(X_i) = \int_0^{X_i} \frac{(F_s x - F_s X_i + F_a Y_i)^2}{2EI_z} dx \qquad (2)$$

$$U_b(X) = \int_0^X \frac{(F_s x)^2}{2EI_z} dx + 2 \cdot (-F_s X + F_a Y) \odot \int_0^X \frac{F_s x}{2EI_z} dx + (-F_s X + F_a Y)^2 \odot \int_0^X \frac{1}{2EI_z} dx \qquad (3)$$

Where $U_b$ is the strain energy due to bending stress, $X, Y$ are the Cartesian coordinates of the tooth profile, $F_a, F_s$ are, respectively, the axial and shear forces, and $EI_z$ is the flexural rigidity of the tooth. The sign $\odot$ denotes the scalar multiplication operator.

2. Efficient calculation of the variable stiffness matrix in cycle $K(cyc)$ – the stiffness matrix is dependent in the gear mesh stiffness (gms) multiplied by a geometrical coefficient, as shown in Equation 4. It is assumed that the gms is uniformly distributed along the z-axis, representing the tooth width. To avoid iterating over both the cycle and the z-axis, a scalar multiplication approach is employed, as demonstrated in Equation 5. By pre-calculating the expectation matrix $\mathbb{E}(z)$ for the z-dependent terms analytically, the computational complexity is reduced to $O(N)$ from the original $O(N \times M)$ complexity, where N and M are the number of cycle points and the number of points along the z-axis, respectively. Similar procedure is applied to the external forces vector $F_{ex}(t)$.

$$K(cyc_i) = K_{const} + \frac{1}{M} \sum_{j=1}^M geom_{i,j} \cdot gms(cyc_i) \qquad (4)$$

$$K(cyc) = K_{const} + [geom_{z_{dep}} \odot \mathbb{E}(z) + geom_{z\_indep}] \odot gms(cyc) \qquad (5)$$

Where $K_{const}$ is the cycle-independent constant structural stiffness. $geom_{z\_dep}$ and $geom_{z\_indep}$ are, respectively, the z-dependent and z-independent geometric coefficients of the gms. The z-dependent geometry is multiplied by the expectation matrix $\mathbb{E}(z)$. For more detailed information, please refer to the user manual [40].

3. Efficient Newton-Rapshon iterations in the numerical solution – Newton-Raphson iterations are employed in the numerical solution of the model, aiming to refine the Newmark method. The Newton-Raphson method involves the inversion of the Jacobian matrix, which is equivalent to the time-varying stiffness matrix $K$ in the model. It is important to note that the length of the time vector is significantly larger than the length of the cycle vector. To optimize computation, instead of calculating the inverse $K(t_i)$ matrix for each time step, the inverse $K(cyc)$ matrix is calculated once for an entire cycle. Then, the corresponding index in the cycle that corresponds to the current time step is selected. This approach is feasible due to the cyclic nature of the stiffness matrix in time.

The optimization process results are presented in Figure 3, which compares the computational time of the basic model outlined in Ref. [30] with the improved model. This comparison is conducted for both a healthy gear and various faulty cases. The figure demonstrates that the improved model successfully decreased the computational time by at least one order of magnitude, and in certain fault scenarios, the reduction reached two orders of magnitude. As previously highlighted, this reduction in computational resources holds significant importance for data-driven technologies.

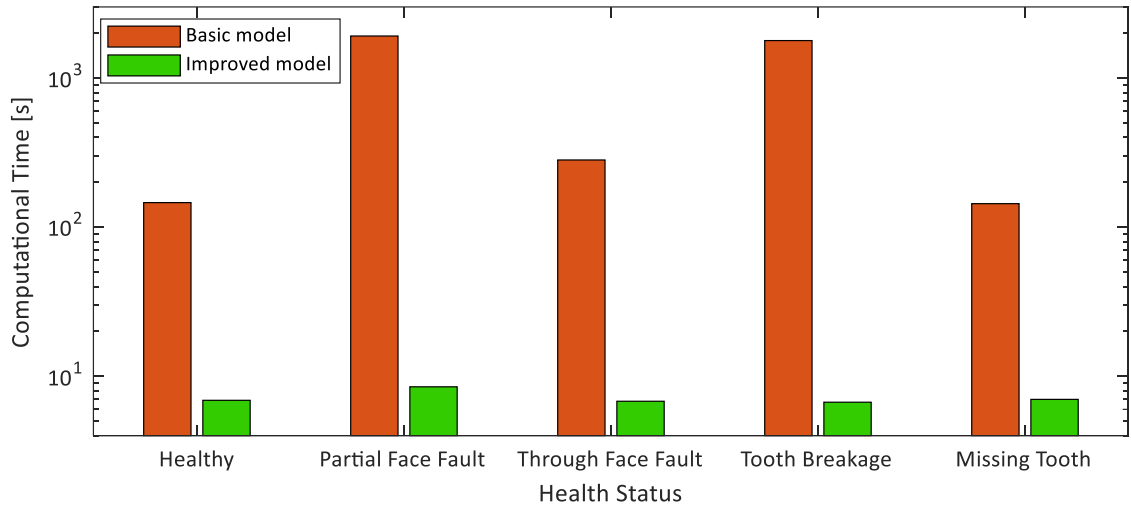

*Figure 3 - Comparative analysis of computational time*

## 5. Simulation enhancement:

Every model is underlain by a set of assumptions that results in differences between the simulated and real signals [36]. Considering every factor in the gear system, such as backlash, lubrication, and surface roughness, is a highly intricate task. Even if the simulated component is perfectly modeled, the transmission path between the component and the sensor creates a gap between the simulation and reality [44]. These differences are inevitable and typically do not hinder the extraction of qualitative insights from the model. However, for most of the machine learning-based algorithms and digital-twins, or certain signal processing algorithms, the requirement for a good resemblance between the simulated and measured signals is crucial. Therefore, in the present study, a novel hybrid approach is proposed to address these discrepancies by enhancing the simulated signal with measured data. This enhanced simulation can serve as training datasets for learning algorithms and potentially lay the groundwork for digital twins.

A block diagram describing the algorithm of the simulation enhancement is presented in Figure 4. Three enhancements are applied to the difference signal [33,45] of the synchronous average [46,47].

- Signal duration modification – the signal is either stretched or shrunk, depending on the desired width ratio. This modification emulates the dispersion of the signal caused by transmission path effects and other factors. When stretching the signal, a modified segment is selected and interpolated to match the original length. On the other hand, when shrinking the signal, it is padded with zeros and interpolated to match the original length.
- Fault-to-harmonics ratio modification – the modified signal ($\text{diff}_{\text{mod}}$) is obtained by taking a linear combination, scaled by $\alpha$, of the original difference signal (diff) and the difference signal corresponding to the healthy state ($\text{diff}_{\text{healthy}}$), as shown in Equation 6. Local tooth faults are expressed by amplified and impulsive response in the difference signal. Hence, this modification fine-tunes the impulsive response generated by the fault.

$$\text{diff}_{\text{mod}} = \alpha \cdot \text{diff} + (1 - \alpha) \cdot \text{diff}_{\text{healthy}} \tag{6}$$

- Signal-to-noise (SNR) modification – the sensor inevitably introduces noise to the signal depending on various factors, including temperature, sensor type, and more. Therefore, a white noise matrix, multiplied by an input factor, is added to the original signal. This not only helps reduce discrepancies with the measured signals but also serves as an augmentation technique, as multiple noisy signals are generated from a single simulation.

The modified simulated signals and the measured signal are normalized by the mean rms of the healthy state prior to feature extraction in order to overcome the attenuation caused by the transmission path effects. The extracted features may include the statistical moments of the difference signal and its envelope, such as rms, skewness, and kurtosis [48,49]. The best set of parameters is determined by selecting the combination that yields the lowest mean absolute errors across the entire dataset. The error is calculated by taking the absolute difference between the condition indicators (CIs) of the enhanced simulation and the measured data, divided by the standard deviation ($\sigma$) of the CIs of the experimental signals, as shown in Equation 7. CIs are computed using the logarithm of extracted features to assign equal importance to all cases.

$$\text{error}_{i,j} = \frac{\left|\text{CI}_{\text{exp}_{i,j}} - \text{CI}_{\text{sim}_{i,j}}\right|}{\hat{\sigma}_{\text{exp}_j}} \tag{7}$$

Where $\text{CI}_{\text{exp}_{i,j}}$ and $\text{CI}_{\text{sim}_{i,j}}$ represent the $j^{\text{th}}$ condition indicator of the $i^{\text{th}}$ signal in the experimental and simulated data, respectively. $\hat{\sigma}_{\text{exp}_j}$ denotes the standard deviation of the $j^{\text{th}}$ CI across all experimental signals.

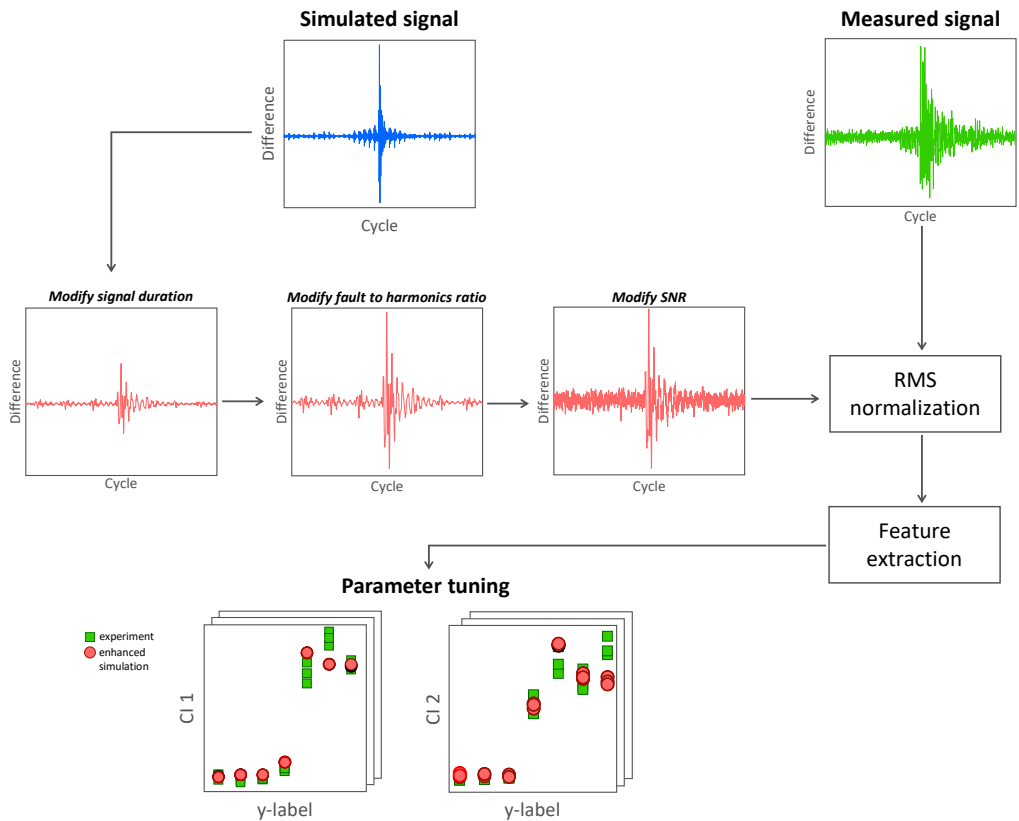

*Figure 4 - Simulation enhancement methodology*

Three experiments were conducted, each employing a test apparatus consisting of an open spur gear transmission. The setup involves a driving motor connected to the input pinion, and a loading source applying torsional torque to the output gear. Additionally, a piezoelectric accelerometer is mounted on the support brackets of the transmission, and a magnetic pickup tachometer is positioned near the shaft, as illustrated in Figure 5 for all experiments. Additional parameters of the experimental setup are outlined in Table 2.

The measured signals underwent angular resampling and segmentation into consecutive intervals. Synchronous average signals were then computed from these segments. Subsequently, the difference signal was calculated, aiming to filter out the gear mesh harmonics and the close pairs of sidebands associated with amplitude modulation phenomena [42] from the synchronous average signal. The model generates vibration signals maintaining the same specifications and operational conditions as the experiment, and difference signals are subsequently calculated.

Various combinations of width ratio, fault-to-harmonics ratio, and SNR were tested. In the examined experiments, optimal parameters were determined by comparing the rms and kurtosis of the difference signal. The combination that resulted in the lowest mean absolute error was selected for enhancement. The optimal parameter values for each experiment are summarized in Table 3.

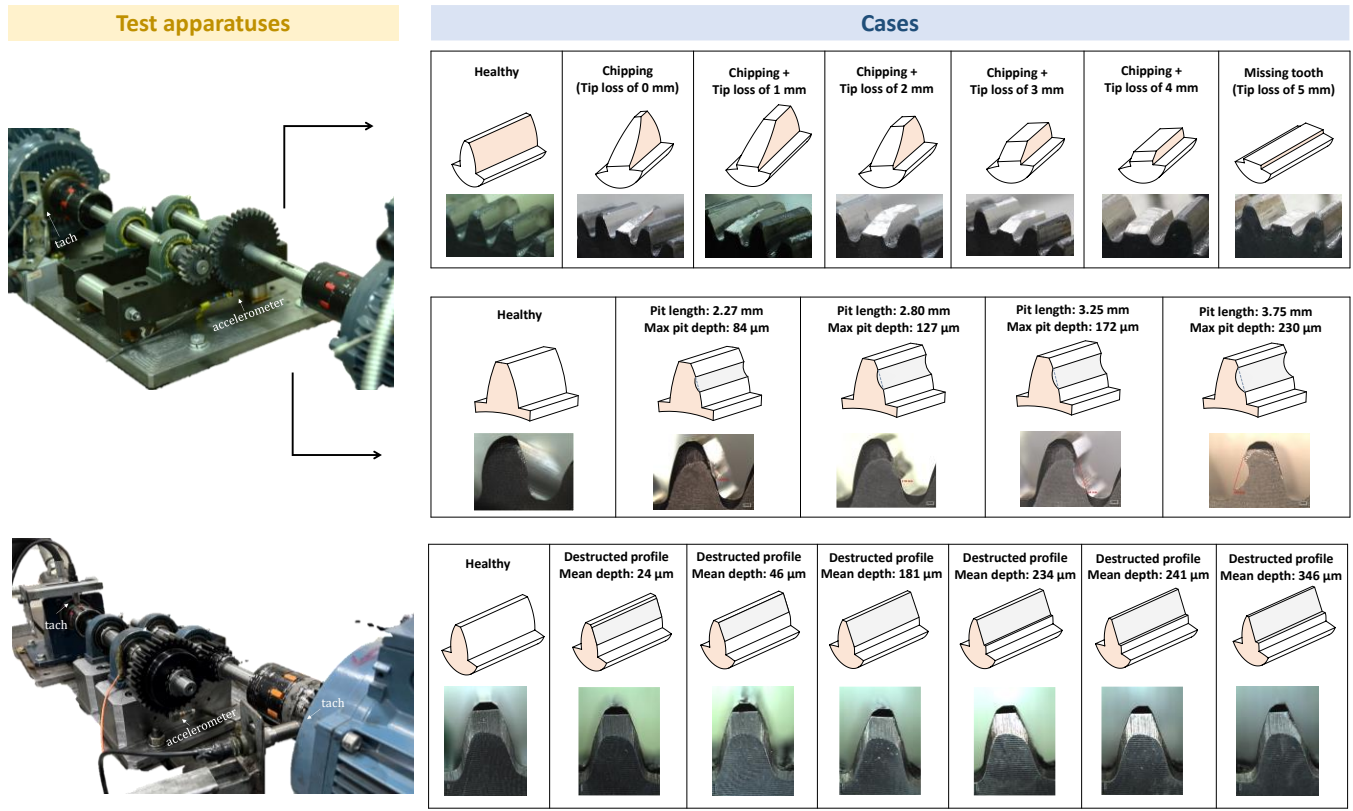

*Figure 5 - The test apparatus: (a) - picture of the experimental system, (b) - sketches and pictures of all health statuses*

Table 2 – Experimental setups

| Name | Tooth Breakage | Pitting | Involute Destruction |
|---|---|---|---|
| Module [mm] | 3 | 3 | 3 |
| Transmission ratio ($z_g : z_p$) | (38 : 17) | (38 : 17) | (35 : 18) |
| Surface roughness | DIN7 | DIN7 | DIN8 |
| Input speed [Hz] | 40 | 40 | 45 |
| Output load [Nm] | 10 | 10 | 15 |
| Cases (healthy + faults) | 7 | 5 | 7 |
| Sampling rate [kS/s] | 25 | 25 | 50 |
| Time duration [s] | 60 | 60 | 60 |
| Number of records | 36 | 30 | 56 |

Table 3 – Calculated optimal parameter values

| Name | Tooth Breakage | Pitting | Involute Destruction |
|---|---|---|---|
| signal width ratio | 0.182 | 0.155 | 0.143 |
| fault-to-harmonics ratio | 2.25 | 1.91 | 1.93 |
| noise level | 1.585 | 1.995 | 1.22 |

The results presented in Figure 6 illustrate a comparison of the condition indicators obtained from the experiments, the original simulation, and the enhanced simulation. Additionally, a representative comparison of the difference signal between the experiment and the enhanced simulation is provided for each experiment. It is evident that there is a strong correspondence between the normalized rms values in all the experiments and the original simulations. Conversely, noticeable differences are observed in the kurtosis values. These two features effectively capture the shape and energy levels of the signal. The enhancement process focused mainly on reducing the discrepancies in the kurtosis values while maintaining a consistent trend in the rms values. Consequently, it is crucial to monitor both of these CIs. By fine-tuning the simulation parameters to minimize the error on the CIs, the enhanced simulation signal closely resembles the measured signal. The successful enhancement process enables the use of the hybrid model to generate a substantial volume of signals that accurately represent the inspected apparatuses. This capability is particularly advantageous in applications such as digital twins, for example.

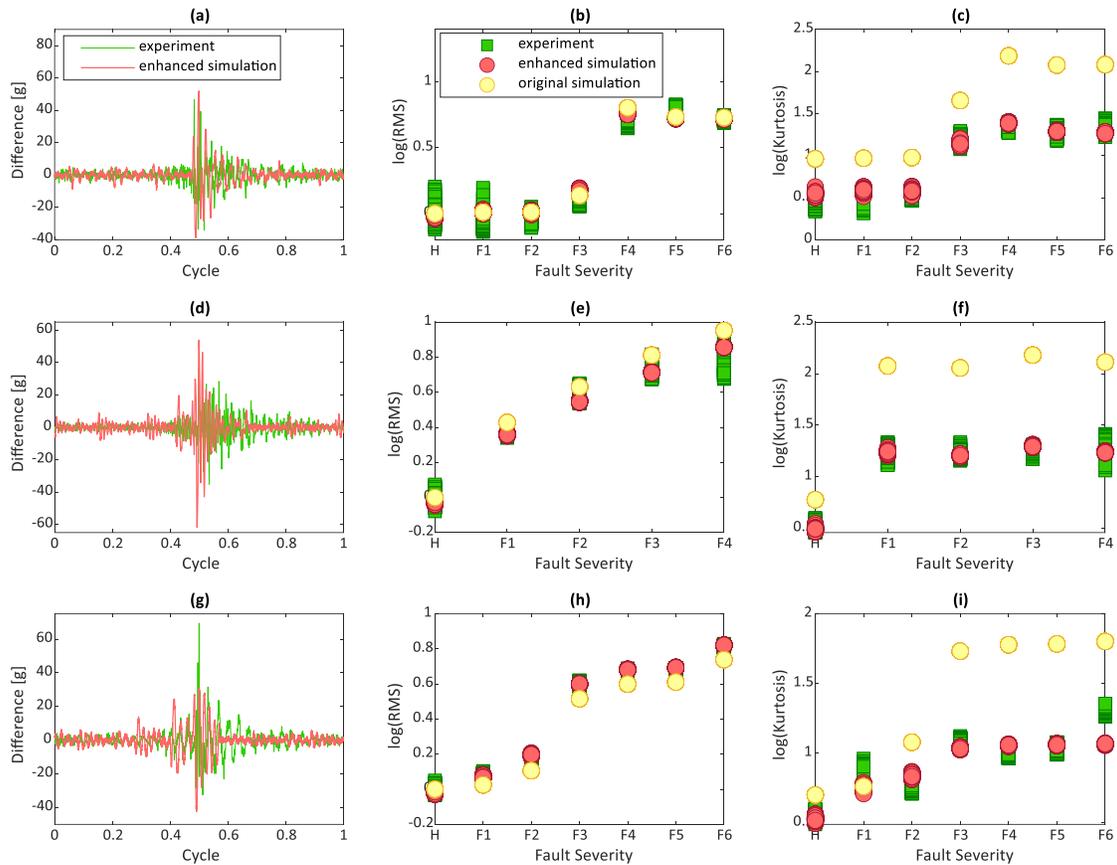

*Figure 6 - Simulation enhancement results: Left column - comparison of the difference signal between the experiment and the enhanced simulations. Middle column - comparison of the rms against tip loss. Right column - comparison of the kurtosis against tip loss. (a-c) Tooth breakage experiment, (d-f) Pitting experiment, (g-i) Involute destruction.*

# 6. Conclusions:

In this study a novel hybrid model that combines realistic dynamic model and a simulation enhancement process is suggested. The dynamic model is made publicly available with a user-friendly application programming interface and a detailed user manual which meticulously explains all its functions. Unlike former work, great attention was invested on the computational running time by implementing new algorithmic processes to improve efficiency. To address the differences between simulated and real signals a novel simulation enhancement process was developed. This process minimizes discrepancies between the features of simulated and real signals by modifying the signal duration, fault-to-harmonics ratio, and signal-to-noise ratio.

The effectiveness of the novel hybrid approach was demonstrated on three different experiments and fault types. The hybrid model holds great potential in contributing to physics-based, machine learning, and digital twin algorithms. Furthermore, the availability of the hybrid model and its implementation to the public can facilitate future development of newer hybrid and dynamic models. To the best of our knowledge, this is the first study to generously share a complete dynamic model along with comprehensive guidance for public access.

**Acknowledgement**

Omri Matania is supported by the Adams Fellowships Program of the Israel Academy of Sciences and Humanities

**Declaration of Generative AI and AI-assisted Technologies in the Writing Process**

During the preparation of this work the authors used ChatGPT 3.5 exclusively to improve readability and language. After using this tool, the authors meticulously reviewed and edited the content as needed and take full responsibility for the content of the publication.